\documentclass[a4paper,10pt]{article}

\usepackage[top=2cm, bottom=2cm, left=1cm, right=1cm]{geometry}

\usepackage{authblk}

\usepackage{color}
\usepackage{amsmath}
\usepackage{amsfonts}

\usepackage{comment}

\newtheorem{lm}{Lemma}
\newtheorem{lem}{Lemma}
\newtheorem{Th}{Theorem}

\newtheorem{rem}{Remark}
\newtheorem{prop}{Proposition}

\newcommand{\RR}{\mathbb{R}}

\newcommand{\V}{\ensuremath{\vec v}}
\newcommand{\vxi}{\vec{\xi}}
\newcommand{\veta}{\vec{\eta}}

\def\<#1>{\langle#1\rangle}

\newcommand{\Xint}[1]{\mathchoice
{\XXint\displaystyle\textstyle{#1}}%
{\XXint\textstyle\scriptstyle{#1}}%
{\XXint\scriptstyle\scriptscriptstyle{#1}}%
{\XXint\scriptscriptstyle\scriptscriptstyle{#1}}%
\!\int}
\newcommand{\XXint}[3]{\setbox0=\hbox{$#1{#2#3}{\int}$ }
\vcenter{\hbox{$#2#3$ }}\kern-.6\wd0}
\newcommand{\dashint}{\Xint-}

\begin{document}

\title{Equilibration in the Kac Model using the GTW Metric $d_2$}
\author[1]{H. Tossounian}
\affil[1]{School of Mathematics, Georgia Institute of Technology, Atlanta}

\maketitle

\begin{abstract}
We use the Fourier based Gabetta-Toscani-Wennberg (GTW) metric $d_2$ to study the rate of convergence to equilibrium for the Kac model in $1$ dimension. We take the initial velocity distribution of the particles to be a Borel probability measure $\mu$ on $\RR^n$ that is symmetric in all its variables, has mean $\vec{0}$ and finite second moment. Let $\mu_t(dv)$ denote the Kac-evolved distribution at time $t$, and let $R_\mu$ be the angular average of $\mu$. We give an upper bound to $d_2(\mu_t, R_\mu)$ of the form $\min\{ B e^{-\frac{4 \lambda_1}{n+3}t}, d_2(\mu,R_\mu)\}$, where $\lambda_1 = \frac{n+2}{2(n-1)}$ is the gap of the Kac model in $L^2$ and $B$ depends only on the second moment of $\mu$. We also construct a family of Schwartz probability densities $\{f_0^{(n)}: \RR^n\rightarrow \RR\}$ with finite second moments that shows practically no decrease in $d_2(f_0(t), R_{f_0})$ for time at least $\frac{1}{2\lambda}$ with $\lambda$ the rate of the Kac operator.
We also present a propagation of chaos result for the partially thermostated Kac model in \cite{TV}.
\end{abstract}

\section{Introduction}

  In \cite{Kac} Kac introduced a linear $n$ particle model with the goal of deriving the Boltzmann equation with Maxwellian molecules. 
He derived a space homogeneous Boltzmann-type equation using the notion of propagation of chaos, which he called the ``propagation of the Boltzmann property''.
A sequence of densities $\{f_n\in L^1(S^{n-1}(\sqrt{nE})), \sigma) \rightarrow \RR\}_n$ on the spheres $S^{n-1}$ where each $f_n$ invariant under the exchange of the variables
 is called chaotic with limit $h$ if

\[ \lim_{n\rightarrow \infty} \int_{S^{n-1}\left(\sqrt{(nE)}\right)} f_n(v_1, \dots, v_n) \phi(v_1, \dots, v_k) \sigma(dv) = \int_{\RR^k} \prod_{i=1}^k h(v_i)\phi(v) dv^k.\]

\noindent for all $k$ and all $\phi\in L^{\infty}$ that depends only on $v_1, \dots, v_k$. Here $E$ is the average energy per particle and is independent of $n$.

\vspace{0.5cm}

Kac showed for his model that if $\{f_n(t=0,.)\}_n$ is a chaotic sequence with limit $f_0$, then so is the time evolved $\{f_n(t,.)\}_n$ for any time $t \geq 0$ and
the chaotic limit $h(t,v)$ of the $\{f_n(t,.)\}_n$ satisfies the Kac-Boltzmann equation

\begin{equation}\label{eq:KacB}
 \frac{\partial h}{\partial t}(t,v) = \int_{\RR} \dashint_0^{2\pi} \left( h(t,v^\ast) h(t, w^\ast) - h(t,v) h(t,w) \right) \, d\theta\,dw
\end{equation}

with initial condition $h(0,v) = f_0(v)$. Here $v^\ast(\theta)$ and $w^\ast(\theta)$ are given by the equation:

\begin{equation}\label{eq:ast}
  (v^\ast(\theta), w^\ast(\theta)) = (v\cos\theta-w\sin\theta , v\sin\theta+w\cos\theta).
\end{equation}

\vspace{0.5cm}

The dynamical variables in Kac's model are the $1$ dimensional velocities of $n$ identical particles. The particles are assumed to be uniformly distributed in space and only their velocities evolve. Let $\vec{v} = (v_1, \dots, v_n)$ denote the velocities of the particles, and $f(t,\vec{v})$ denote the distribution of the velocities. 
A binary collision takes place at a sequence of random times $\{t_i\}$ with  $\{t_{i+1}-t_i\}$  i.i.d. with law exp$(n \lambda)$, for some parameter $\lambda$ independent of $n$ as follows.
At $t_i$, pair of particles $(k,l)$ is chosen randomly and uniformly among the ${n\choose 2}$ pairs to collide. Let $v_k$ and $v_l$ be their velocities prior to the collision. After the collision their velocities become $v_k^\ast(\theta)$ and
$v_l^\ast(\theta)$ given by equation \eqref{eq:ast} with $v$ and $w$ replaced by $v_k$ and $v_l$, and where $\theta$ is chosen randomly and uniformly in $[0,2\pi]$. These collisions preserve energy.
\vspace{0.5cm}

\noindent We represent the effect of rotating particles $k$ and $l$ on a probability density $f$ by the operator $Q_{i,j}$. $Q_{i,j}$ is given by:

\begin{equation}\label{eq:Qij}
Q_{k,l}f = \dashint_0^{2\pi} f(v_1, \dots, v_{k-1}, v_k^\ast(\theta), \dots, v_l^\ast(\theta), \dots, v_n)\,d\theta,
\end{equation}

\noindent and the collision operator $Q$  is given by $\displaystyle Q={n \choose 2}^{-1} \sum_{i<j} Q_{i,j}$.

\vspace{0.5cm}

    The Fokker-Planck equation of this process is known as the Kac master equation and is given by

\begin{equation}\label{eq:Kac}
    \frac{\partial f(t,\vec{v})}{\partial t} =n \lambda(Q-I) f:= - L f.
\end{equation}

\noindent Here $-L$ is the generator of the Kac process. Kac worked on the sphere $\sum_{i=1}^n v_i^2= n E$ and took the initial distributions to be a symmetric under the exchange of its variables. 
$L^2\left(S^{n-1} (\sqrt{nE})\right)$. This symmetry, which is preserved by the Kac evolution, is the physically interesting case.
The restriction to a sphere is possible because Kac's evolution preserves the energy $v_1^2+ \dots + v_n^2$ and therefore preserves the property of being supported on a sphere too. It is well known (see the introduction of \cite{CCL}) that on each sphere the only stationary solutions are the constants and that the Kac process is ergodic. On $\RR^n$, i.e. when the energy at $t=0$ is not fixed, the equilibria are the radial functions.

\vspace{0.5cm}

In the following, let

\begin{itemize}
  \item $\sigma^r$ ( or $\sigma$ if $r$ is clear from the context) denote the normalized uniform probability measure on $S^{n-1}(r)$ for any $r>0$;
  \item $\vert h\vert_{L^p(r)}^p$ be $\int_{S^{n-1}(r)} \vert h(w)\vert^p \sigma^r(dw)$ for $1\leq p<\infty$,
  \item $\vert h\vert_{L^\infty(r)} = \emph{ess\, sup} \{ \vert h(w)\vert: \vert w \vert=r\}$, and 
  \item $R_h$ denote the angular average of $h$: $R_h(v) = \int_{S^{n-1}(\vert v\vert)} \,h(w) \sigma^r(dw)$. $R_\mu$ 
can be defined similarly for Borel probability measures $\mu$. ($R_h$ was called the radial average of $h$ in \cite{BLV} and \cite{TV}.)
 \item $Q_{i,j}(\theta)$ map $(v_1, \dots, v_n)$ to $(v_1, \dots, v_i\cos\theta-v_j\sin\theta, v_{i+1}, \dots, v_i\sin\theta-v_j\cos\theta, v_{j+1} , \dots, v_n)$.
\end{itemize}

\vspace{0.5cm}

The aim of this paper is to study the Gabetta-Toscani-Wennberg metric $d_2$ in relation with the Kac evolution, and to give a propagation of chaos result for the partially thermostated 
Kac model in \cite{TV}. The speed of approach to equilibrium is one of the central questions in this field. Kac in \cite{Kac} conjectured that there is a spectral gap for the generator of the master equation on $L^2(S^{n-1}(r))$ that is independent of the number of particles. 
Kac's conjecture was proved by Janvresse in \cite{Janvresse} and the gap was computed explicitly in \cite{CCL}, where the authors show if $f: L^2(S^{n-1}(r))\rightarrow \mathbb{R}$ is symmetric in its variables with integral $1$, then the following inequality holds:
 
\begin{equation}\label{eq:l2gap}
\vert\vert e^{- L t}f - 1 \vert\vert_{L^2(r)} \leq e^{-\lambda \frac{n+2}{2(n-1)} t} \vert\vert f - 1 \vert\vert_{L^2(r)}.
\end{equation}


The $L^2$ gap requires time of order $n$ to show fast convergence to equilibrium because the initial norm $\vert\vert f - 1 \vert\vert_{L^2(r)}$ 
can grow exponentially in $n$ if $f=\prod f_1(v_i)/Z$ is a normalized product on $S^{n-1}(r)$.

\vspace{\baselineskip}

The (negative) of the relative entropy $S(f(t)\vert 1)= \int f\ln\left(\frac{f}{1}\right)\,d\sigma^r$ was studied as a distance to equilibrium because it is an extensive quantity. We have $S(f\vert 1) \geq 0$ and $S(f\vert 1)= 0$ if and only if $f =1$ a.e. Villani showed in \cite{Villani} that

\begin{equation}\label{eq:Villani}
 S(f(t)\vert 1) \leq e^{-\frac{2\lambda}{n-1}t} S(f \vert 1),
\end{equation}

\noindent using entropy production techniques. The initial entropy production is defined by $-\left. \frac{1}{S(f(t)\vert 1)}\frac{d}{dt} S(f(t)\vert 1) \right\vert_{t=0+}.$ Einav showed in  \cite{Einav} that the rate in \eqref{eq:Villani} is essentially sharp in the $n$ behavior at $t=0$, disproving Cercignani's conjecture in the context of the Kac model which states that 
there is a positive lower bound on the entropy production that is independent of $n$ for the class of $L^1$ functions with finite entropy and finite second moment
 (see \cite{Cerc2} and Section $6$ in \cite{Villani}.)

\vspace{\baselineskip}

Exponentially fast decay with rate independent of the number of particles was established in \cite{BLV} for the Kac model coupled to a thermostat. In this model, the particles in addition to colliding among themselves, collide at a rate $\eta$ with particles from a Maxwellian thermostat at a fixed temperature $\beta^{-1}$. 
The energy of the system is no longer conserved since the thermostats can pump in or drain out energy from the system. So, in this model, the solution $f(t, .)$ is supported on all of  $\RR^n$. Equilibrium is reached when all the (non-thermostat) particles are independent and have the Gaussian distribution at the same temperature as the thermostat. 

\vspace{0.5cm}

Motivated by the result in \cite{BLV},  Vaidyanathan and I worked in \cite{TV} with the Kac model where we thermostated $m$ of the particles, $m<n$ using a stronger thermostat at temperature $\beta^{-1}$. Let $P_i$ be the operator representing the action of the strong thermostat on the $i^{th}$ particle. $P_i$ is given by

\begin{equation}\label{eq:Pf}
 P_i[f] (v_1, \dots, v_n) =  g_\beta(v_i) \int_{\RR}  f(v_1, \dots, v_{i-1}, w, v_{i+1}, \dots, v_n)\,dw,
\end{equation}

\noindent where $g_\beta(v)$ the Gaussian at temperature $\frac{1}{\beta}$:

\begin{equation}\label{eq:gbeta}
 g_\beta(x)=\sqrt{\frac{\beta}{2}}\,e^{-\frac{\beta}{2} x^2}.
\end{equation}

\noindent The generator of the partially thermostated Kac model in \cite{TV} is given by

\begin{equation}\label{eq:Lnm}
 -L_{n,m} = n \lambda(Q-I) + \eta \sum_{i=1}^m (P_i -I).
\end{equation}

\noindent The minus sign is there to make $L_{n,m}$ positive definite in $L^2(\RR^n)$.

\vspace{\baselineskip}
 A propagation of chaos result for the partially thermostated Kac model will be presented below, where the $f_n$-s are supported on all of $\RR^n$ instead of the only on the spheres $S^{n-1}(\sqrt{nE})$.

\vspace{\baselineskip}

The Fourier based GTW metric $d_2$ was used in \cite{BLTV} to show that the infinite thermostat model in \cite{BLV} can be approximated uniformly in time by the Kac model with a finite reservoir 
having $n+\mathcal{N}$ particles. Here $\mathcal{N}>>n$ and the initial conditions are taken to have the special form $f(\vec{v}) = l_0(v_1,\dots, v_n) \prod_{n+1}^{n+\mathcal{N}} g_\beta(v_i)$. The last $\mathcal{N}$ particles are the reservoir particles. This approximation was proven under a technical finite fourth moment assumption.
 
\vspace{\baselineskip}

Let $\mu$ and $\nu$ be Borel probability measures on $\RR^n$. The GTW metrics $d_\alpha$ are given by

\begin{equation}\label{eq:dalpha}
d_\alpha(\mu,\nu) = \sup_{\xi \neq \vec{0}} \frac{ \vert \hat{\mu}(\xi) - \hat{\nu}(\xi) \vert}{\vert\xi\vert^\alpha}.
\end{equation}

\noindent Here we use the convention that the Fourier transform of $\phi$ is $\hat{\phi}(\xi)= \int_{\RR^n} \phi(v) e^{-2\pi i \xi.v}\,dv$. We will use only $d_2$ even though analogs of Theorems \ref{lm:d2} and \ref{Th:on} are valid for any $d_\alpha$ with $\alpha>0$. 

\vspace{0.5cm}

The GTW metrics $\{d_\alpha\}_{\alpha>0}$ were introduced in \cite{GTW} in the context of the space homogeneous Kac-Boltzmann equation \eqref{eq:KacB} where they helped in showing 
exponentially fast convergence to equilibrium for the initial data with finite $2+\epsilon$ moment for some $\epsilon>0$. $d_1$ and $d_2$ were used in \cite{CLM} to show exponential
 convergence to steady states for the Kac Boltzmann system coupled to multiple Maxwellian thermostats at different temperatures. Similarly, $d_1$ and $d_2$ were used by J. Evans in
 \cite{Evans} to show existence and ergodicity of non-equilibrium steady states in the Kac model coupled to multiple thermostats.

\vspace{0.5cm}

An interesting feature of the $d_2$ metric that we will elaborate in Sections $2$ and $3$ is its intensivity property given in \cite{BLTV}: Let $f_1, \dots, f_n$ and $g_1, \dots, g_n$ be probability densities on $\RR$ with finite 
second moments and $0$ first moment. Then

\begin{equation}\label{eq:chaotic}
 d_2( \prod_{i=1}^n f_i(v_i), \prod_{j=1}^n g_j(v_j)) = \max_{i\leq n} d_2( f_i, g_i).
\end{equation}

\vspace{0.5cm}

   We take our initial distribution $\mu$ to be a Borel probability measure $\mu$ on $\RR^n$. A special case is a density on $S^{n-1}(\sqrt{n E})$. We adapt equation \eqref{eq:Kac} to measures 
and study the Kac-evolved $\mu$, $e^{-tL}\mu$ using the GTW distance $d_2$. In Section \ref{sec:d2v} we give the ``almost'' intensivity properties of the $d_2$ metric. Proposition \ref{prop:d2v} shows that, after time of $O(\ln n)$, a good quantity to compare $d_2(\mu, R_\mu)$ with is $\int \frac{\vert v\vert^2}{n} \mu(dv)$.
While at $t=0$, there are states for which $d_2(\mu,R_\mu)$ is as big as $\int \vert v\vert^2 \mu(dv)$ which is of order $n$. The function $d_2( e^{-tL}\mu, R_\mu)$ is not guaranteed to be differentiable with respect to $t$
due to the supremum taken in the definition of $d_2$. So Cercignani's conjecture cannot be formulated in the same way as in the relative entropy. But one could formulate the following conditional statement:\\

\noindent ``(C) Let $\mu$ be a Borel probability measure on $\RR^n$ with finite second moment and zero first moment. If $d_2(\mu, R_\mu)>0$ and $d_2(e^{-tL}\mu, R_{\mu})$ is differentiable at $t=0$ then $\frac{ \left. \frac{d}{dt}(d_2(e^{-tL}\mu, R_\mu)) \right\vert_{t=0}}{d_2(\mu,R_\mu)}\geq a$ 
for some $a>0$ independent of $\mu$ or $n$." We will disprove this conjecture in Theorem \ref{Th:on}.\\

 In Section \ref{sec:upperbound} we give the first main theorem: 
Theorem \ref{lm:d2}, a convergence result that provides an upper bound for $d_2(e^{-tL}\mu, R_\mu)$ when $\mu$ has zero mean and finite second moment, and is symmetric under the exchange
of its variables. This upper bound has the form $\min\{B e^{-\frac{4 \lambda_1}{n+3}t}, d_2(\mu,R_\mu)\}$ with $B$ depending only on the second moment of $\mu$. 
This shows that $d_2(e^{-tL}\mu, R_\mu)$ goes to zero. It is curious that the proof uses the $L^2$ gap of the Kac evolution in equation \eqref{eq:l2gap} in an unexpected context. We show in Proposition \ref{prop:d2v} that our bound has the correct order of magnitude at $t=0$. This upper bound gives decay after time of order $n$, in agreement with the 
upper bounds using the $L^2$ and relative entropy metrics. Next,  in Section \ref{sec:f0}, we use the $L^\infty$ nature of the $d_2$ metric to construct a family of functions $f_n \in L^1(\RR^n)$ having 
$O(t^{n-1})$ decay in $d_2$ when $0 \leq t \leq 1/(2\lambda)$. This disproves the Cercignani-type conjecture (C) for the Kac evolution in the $d_2$ metric. We give the construction in Theorem \ref{Th:on}. 
In Section \ref{sec:chaos}, we give a propagation of chaos result for the partially thermostated Kac model in \cite{TV} by adapting McKean's proof of propagation of chaos for the regular Kac model in \cite{McKean}.
In Section \ref{sec:conc} we give some concluding remarks. All the results are stated in Section \ref{sec:results}.

\section{Results}\label{sec:results}

We first give Proposition \ref{prop:d2v}that generalizes equation \eqref{eq:chaotic}. It says that  $\int \frac{\vert v\vert^2}{n} \,\mu(dv)$  essentially gives the order of magnitude of the $d_2(e^{-tL}\mu,R_\mu)$, distance between a measure and its angular average.

\begin{prop}{($d_2$-energy comparison)}\label{prop:d2v} Let $\mu$ and $\nu$ be Borel probability measures on $\RR^n$ with $n\geq 2$.  Let $\int \vec{v} \mu(dv)=\vec{0},  \int \vec{v}\nu(dv)=\vec{0}$, and $\int \vert v\vert^2 (\mu(dv)+\nu(dv)) <\infty$. Let $-L = n(I-Q)$  $(\lambda=1)$ be the generator of 
the Kac evolution ($\lambda=1$). Then

\begin{align} 
 d_2( e^{-tL}\mu, R_\mu) &\leq \frac{(2\pi)^2}{2}\left[\left(2 - e^{- \frac{n}{n-1} t}\right)\int \frac{\vert v\vert^2}{n}\mu(dv) + \right. \nonumber \\
                 &\left.  e^{-\frac{n}{n-1} t}\max_i \int v_i^2 \mu(dv) + (n-1) e^{-\frac{4n-6}{n-1}t}\max_{i \neq j}\vert \int v_i\,v_j\,\mu(dv)\vert \right], \label{eq:symmetric}\\
 d_2(e^{-tL} \mu, e^{-tL}\nu) & \leq  \frac{(2\pi)^2}{2} ( (n-1) e^{-t} + 1) \int_{\RR^n}\left\vert \mu(dv)- \nu(dv)\right\vert \frac{\vert v \vert^2}{n}. \label{eq:propd2vcompare}
\end{align}

\end{prop}

\begin{rem}
If $\mu$ has mean $\vec{m} \neq \vec{0}$. Then $d_2(\mu, R_{\mu})=\infty$ because the angular average $R_{\mu}$ has mean $\vec{0}$. One way around this is to use a centered GTW distance $d'_2$ as in \cite{CLM} and \cite{Evans}. This handles the $\frac{1}{\vert \xi\vert}$ divergence as $\vec{\xi}\rightarrow \vec{0}$ in the definition of $d_2$. We will omit this case. 
\end{rem}

With the help of this proposition, the statement of  Theorem \ref{lm:d2} becomes more natural. 

\begin{Th}\label{lm:d2} Let $\mu$ be a Borel probability measure on $\RR^n$ that is invariant under permutation of coordinates. Let 
\\$ \int \vert v\vert^2 \mu(dv) <\infty$ and $\int \vec{v} \mu(dv) = \vec{0}$. And let $\lambda$ in \eqref{eq:Kac} be $1$. Then

\begin{equation}\label{eq:d2}
 d_2(e^{-tL} \mu, R_\mu) \leq \min\left\{ K\left( e^{-\frac{4 \lambda_1}{n-1}t} \right) \left[2 \int v_1^2 \vert \mu\vert (dv) + (n-1) e^{-\frac{4n-6}{n-1} t} \left\vert \int_{\RR^n} v_1 v_2  \mu(dv) \right\vert\right], d_2( \mu, R_\mu)\right\}.
\end{equation}

\noindent $K= 6.64 (2\pi)^2$ and $\lambda_1$ is the gap in \eqref{eq:l2gap}.

\end{Th}

Theorem \ref{lm:d2} implies that $d_2(e^{-tL}\mu, R_{\mu}) \leq K (n e^{-t} + 1) \int \frac{\vert v\vert^2}{n}\, \mu(dv) \left( e^{-\frac{4 \lambda_1}{n+3}t} \right)$ for all $t$, and that if
$\mu$ has zero correlations between the $v_i$ (e.g. $\mu=\prod_{i}\mu_0(dv_i)$ and $\mu_0$ centered at $0$), then $(n e^{-t} + 1)$ can be replaced by $1$. 
The important information in this theorem is the exponential rate of decay $\frac{4\lambda_1}{n+3}$ for large time. The constant $K$ is not optimal at $t=0$.
It would be desirable to have a bound of the form $d_2( e^{-tL} \mu, R_\mu) \leq 1 e^{-c t/n} d_2(\mu, R_\mu)$. But Theorem $\ref{Th:on}$ implies that no such bound
 exists at least on $[0,1/2]$ even if  $\mu$ has a Schwartz density with respect to the Lebesgue measure. Theorem \ref{Th:on} also implies that, for some Schwartz densities $f$, 
$\left.\frac{d}{dt}  d_2(e^{-tL}f, R_f)\right\vert_{t=0}$ exists and equals $0$.  The conjecture that `` the best constant $K_{\mbox{best}}$ in equation \eqref{eq:d2} satisfies 

\begin{equation}\label{eq:conjecture}
   K_{\mbox{best}}(n) \geq H\left(\int v_1^2 \mu(dv), \int v_1\,v_2 \mu(dv)\right) \left(1 + \frac{c}{n}\right)
\end{equation}

\noindent for some optimal $H$ that is at most linear in its arguments.'' is consistent with Proposition \ref{prop:d2v} and Theorem $2$ because there is decay in equation \eqref{eq:conjecture} only after time of order $1$ (since $( 1 + \frac{c}{n}) e^{-\frac{t}{n}} \leq 1$ when $t \geq n \ln(1 + \frac{c}{n}) \approx c$).

\begin{Th}\label{Th:on} Let $n \geq 2$ and let $L$ be as in equation \eqref{eq:Kac} with $\lambda=1$. There is a Schwartz probability density $f_n$ on $ \RR^n$ that satisfies 

\begin{equation}\label{eq:slowdecay}
    d_2( e^{-tL} f_0, R_{f_0}) \geq \max\left\{d_2(f_0, R_{f_0}) \left( 1 - \frac{e}{n}(2t)^{n-1} \right), 0\right\} \, \mbox{ for all $t \geq 0$}.
\end{equation}

\end{Th}

The lower bound in Theorem \ref{Th:on} endures for $t \in [0,1/2]$. We will give the $f_n$ explicitly in Lemma \ref{lm:f01} up to two parameters $\bar{A}(n)$ and $B(n)$ that are shown to be finite but are not computed. The functions $f_n$ will be perturbations of the Gaussians $\prod_{i=1}^n\Gamma_{\alpha(n)}(v_i)$
 at high temperature by Schwartz functions that have small $L^1$ norms.

\vspace{0.5cm}

Finally, we give the propagation of chaos result for the partially thermostated Kac model in \cite{TV}. This result is independent of the previous Theorems. As mentioned in the introduction, the energy of the system of particles is no longer conserved. Thus our functions will be in $L^1(\RR^n)$ for various $n$ instead of $L^1(S^{n-1}(\sqrt{nE}))$. Let $n_0, m_0<n_0$ be such that $\alpha = \frac{m_0}{n_0}$ is the fraction of particles that are thermostated. Let $L_{m,n}$ be given by \eqref{eq:Lnm}. Then we have the following theorem.

\begin{Th}\label{Th:chaos}{Propagation of Chaos for the Partially thermostated Kac Model}\\

Let $A= \{ i: i\geq 1, \mbox{ and } (i \mod n_0) \in \{ 1,2, \dots, m_0\}\,\}$ and $B = \mathbb{N}- A$. Let $\{ f_k \in L^1(\RR^{ k n_0})\}_{k=1}^\infty$ be a family of probability distributions that are symmetric under the exchange of particles with indices in $A$ and under the exchange of particles with indices in $B$. If

\[    \lim_{k\rightarrow \infty} \int_{\RR^{k n_0}} f_{k}(v_1, \dots, v_{k n_0})\phi(v_1, \dots, v_l) \,dv =  \int_{\RR^{k n_0}} \prod_{i\in A, i \leq l} \bar{f}_0(v_i) \prod_{j \in B, j\leq l} \bar{\bar{f}}_0(v_j)\, \phi(v_1, \dots, v_l)dv,\]

\noindent for every $\phi$ in $L^\infty(\RR^l)$, then 

\[    \lim_{k\rightarrow \infty} \int_{\RR^{k n_0}} e^{-tL_{km_0, k n_0}} [f_{k}](v_1, \dots, v_{k n_0})\phi(v_1, \dots, v_l) \,dv = \int_{\RR^{k n_0}} \prod_{i\in A, i \leq l} \bar{f}(t,v_i) \prod_{j \in B, j\leq l} \bar{\bar{f}}(t,v_j)\, \phi(v_1, \dots, v_l)\, dv,  \]

\noindent for every $\phi$ in $L^\infty(\RR^l)$ where $(\bar{f}, \bar{\bar{f}})$ satisfy the following system of Boltzmann-Kac equations:

\begin{equation}\label{BoltzmannKacPartial}
 \left\{ \begin{array}{cl} \frac{\partial \bar{f}}{\partial t} (t,v) & = 2\lambda \left[  \int_{\RR}\dashint_{0}^{2\pi} \bar{f}(t,v^\ast) (\alpha \bar{f}(t,w^\ast) + (1-\alpha) \bar{\bar{f}}(t,w^\ast))\, d\theta\,dw - \bar{f}(t,v)\right]+\eta (P_1 - I)\bar{f} \\
 \frac{\partial \bar{\bar{f}}}{\partial t}(t,v)& = 2\lambda \left[ \int_{\RR}\dashint_{0}^{2\pi} \bar{\bar{f}}(t, v^\ast) (\alpha \bar{f}(t, w^\ast) + (1-\alpha) \bar{\bar{f}}(t,w^\ast))\, d\theta\,dw - \bar{\bar{f}}(t,v)\right] 
  \end{array}\right. ,
\end{equation}

\noindent together with the initial conditions $(\bar{f}(t=0), \bar{\bar{f}}(t=0)) = (\bar{f}_0, \bar{\bar{f}}_0)$. 

\end{Th}

\noindent This roughly says that a given particle collides with a thermostated particle a fraction $\alpha$ of the time, and with non-thermostated particles: the fraction $1-\alpha$ of the time.

\section{Proof of Proposition \ref{prop:d2v}}\label{sec:d2v}

The proof of Proposition \ref{prop:d2v} relies on the action of the Kac evolution on quadratic polynomials. The following lemma says
that after time of order $\ln(n)$, $(v.\xi)^2$ is effectively $\frac{\vert v\vert^2}{n} \vert \xi\vert^2$.

\begin{lm}{(Kac action on Quadratic Polynomials)}\label{lm:quad} Let $n \geq 2$ and let $L$ be as the operator in $\eqref{eq:Kac}$ with $\lambda =1$. For any $v, \xi \in  \RR^n$, we have

\begin{equation}\label{eq:E1}
 e^{-tL}(v.\xi)^2 = \left(1 - e^{- \frac{n}{n-1} t}\right) \frac{\vert v\vert^2 \vert \xi\vert^2}{n} + e^{-\frac{n}{n-1}t} \sum_{i=1}^n \xi_i^2 v_i^2  +    e^{- \frac{4 n -6}{n-1} t} \sum_{i \neq j} \xi_i \xi_j v_i v_j 
\end{equation}

It follows that for all $n\geq 2$ and $t \geq 0$ we have

\begin{equation}\label{eq:E2}
\left\vert e^{-tL} (v.\xi)^2 - \frac{\vert v\vert^2 \vert \xi\vert^2}{n}\right\vert \leq e^{-t}\left(1- \frac{1}{n}\right) \vert v\vert^2 \vert \xi \vert^2.\end{equation}

\end{lm}

\noindent \textbf{Proof} ( of Lemma \ref{lm:quad}) We will look at the action of $Q$ on $v_1 v_2$ and on $v_1^2$ separately. First,

\[Q_{i,j} v_1 v_2 = \left\{ \begin{array}{cc} 0, & \{i,j\} \cap \{1, 2\} \neq \phi\\ v_1 v_2, & \mbox{ otherwise} \end{array}\right.\]

It follows that  $e^{-tL} v_1 v_2 = e^{-n( 1 - \frac{ {n-2\choose2}}{{n \choose 2}} ) t} v_1 v_2$ . Similarly, $Q v_1^2 = ( 1 - \frac{1}{n-1}) v_1^2 + \frac{1}{n-1} \frac{\vert v\vert^2}{n}$. Thus

\[ n(Q-I) \left(  \begin{array}{c} v_1^2 \\ \frac{\vert v\vert^2}{n} \end{array}\right) = \left( \begin{array}{cc} -\frac{n}{n-1} & \frac{n}{n-1} \\ 0 & 0 \end{array}\right)
   \left(  \begin{array}{c} v_1^2 \\ \frac{\vert v\vert^2}{n} \end{array}\right)\]

And since $\exp \left(t\left( \begin{array}{cc} -\frac{n}{n-1} & \frac{n}{n-1} \\ 0 & 0 \end{array}\right)\right) = 
 \left( \begin{array}{cc} e^{-t\frac{n}{n-1}} & 1 - e^{-t\frac{n}{n-1}} \\ 0 & 1 \end{array}\right)$, we obtain:

\[\displaystyle e^{-tL} v_1^2 = e^{-\frac{n}{n-1} t} v_1^2 + ( 1 - e^{-\frac{n}{n-1} t}) \frac{\vert v\vert^2}{n}.\]

\noindent  From these two identities equation \eqref{eq:E1} follows.

Next we prove equation \eqref{eq:E2}. Let $a = n\left( 1 - \frac{{n-2 \choose 2}}{{n \choose 2}}\right) = \frac{4 n - 6}{n-1}$ and let $b= \frac{n}{n-1}$. We have $a \geq 2 b$ when $n \geq 2$. The right-hand side of \eqref{eq:E2} can be written as 

\[\displaystyle e^{-a t}  (v.\xi)^2 + (e^{- b t} -  e^{- a t}  ) \sum_{i=1}^n \xi_i^2 v_i^2 - \frac{\vert \xi\vert^2 \vert v\vert^2}{n} e^{-bt}.\]

\noindent  This is bounded above by $e^{-b t} \left( 1 - \frac{1}{n} \right) \vert v\vert^2 \vert \xi\vert^2$ because $\xi_i^2 \leq \vert \xi\vert^2$.  Similarly, $\displaystyle  e^{-tL} (v.\xi)^2 - \frac{\vert v\vert^2 \vert \xi\vert^2}{n} \geq - \frac{\vert v\vert^2 \vert \xi\vert^2}{n}$. Taking absolute values and using the observation that $1-\frac{1}{n} \geq \frac{1}{n}$ completes the proof.
$\hfill \square$

\vspace{0.5cm}

We are now ready to prove Proposition \ref{prop:d2v}.\\

\noindent \textbf{Proof} (of Proposition \ref{prop:d2v})  We start with the definition of $d_2( e^{-tL}\mu, R_\mu)$.

\begin{eqnarray*}
d_2( e^{-tL}\mu, R_\mu) & = & \sup_{\xi \neq 0}\frac{1}{\vert \xi\vert ^2}\left\vert \int_{\RR^n} \left( e^{-tL}\mu(dv)-R_\mu\right) e^{-2\pi i v.\xi} \right\vert \\ & = &  \sup_{\xi \neq 0} \frac{1}{\vert \xi\vert ^2}\left\vert \int_{\RR^n}  e^{-tL}\mu(dv) \left( e^{-2\pi i v.\xi} -R_{ e^{-2\pi i v.\xi}}(v) \right)\right\vert \\ 
\end{eqnarray*}

\noindent Here we used the self-adjointness of taking the angular average and the fact that $e^{-tL}\mu$ and $\mu$ have the same angular average. Here  $R_{e^{-2\pi i v.\xi}}$ is the angular average of $\exp(-2\pi i v.\xi)$ which we study next. For brevity, let $R$ denote $R_{e^{-2\pi i v.\xi}}$. Then $R$ is also the angular average of $\cos(2\pi v.\xi)$ and we have

\begin{eqnarray*}
 R(v) & = & \dashint_{ \vert y\vert = \vert v\vert} \cos(2\pi y_n \vert \xi\vert)\,dy = \left\vert S^{n-1}\right\vert^{-1}\int_{ S^{n-1}} \cos( 2\pi \vert v\vert \vert \xi\vert \cos\theta_1 )\,d\sigma^n\\
        & = &  \frac{ \vert S^{n-2}\vert}{\vert S^{n-1}\vert} \int_{\theta_1=0}^\pi \cos(2\pi\vert v\vert \,\vert \xi\vert \cos\theta_1)\sin(\theta_1)^{n-2}\,d\theta_1\\
        & = & \frac{  \int_{\theta_1=0}^\pi \cos(2\pi \vert v\vert \, \vert \xi\vert \cos\theta_1)\sin(\theta_1)^{n-2}\,d\theta_1}{ \int_{\theta_1=0}^\pi \sin(\theta_1)^{n-2}\,d\theta_1}.
\end{eqnarray*}

We won't need this fact, but $R(v) =_0F_1(\frac{n}{2}, - \frac{ (2\pi \vert v\vert \vert \xi\vert)^2}{4})$. Here $_0F_1$ is the hypergeometric function given by $_0F_1(a,x) = 1+ \sum_{k=1}^\infty \frac{1}{a(1+a)(2+a) \dots (k-1+a)}\frac{x^k}{k!}$.
Going back to $d_2( e^{-tL}\mu, R_\mu)$, we can use the fact that $\int v_i  \mu(dv) = 0$ for every index $i$ to write 

\begin{eqnarray*}
d_2( e^{-tL}\mu, R_\mu)& = & \sup_{\xi \neq 0} \frac{1}{\vert \xi\vert ^2}\left\vert \int_{\RR^n}  e^{-tL}\mu(dv) \left( e^{-2\pi i v.\xi} -1+ 2\pi i v.\xi + 1 - R_{ e^{-2\pi i v.\xi}}(v) \right)\right\vert,\\
	\mbox{thus } d_2(	e^{-tL}\mu, R_\mu )& \leq & \sup_{\xi \neq 0} \int_{\RR^n}  e^{-tL}\mu(dv) \frac{\left\vert e^{-2\pi i v.\xi} -1  + 2\pi i v.\xi \right\vert}{\vert \xi\vert^2} + \int_{\RR^n} e^{-tL}\mu (dv) \frac{1 - R(v)}{\vert \xi\vert^2}\,dv.\\
\end{eqnarray*}

\vspace{0.5cm}

Taylor's theorem gives  $\vert e^{ix} - 1- ix\vert \leq \frac{1}{2} x^2$ for all $x \in \RR$. We thus have

\begin{equation}\label{eq:d2a1}
\sup_{\xi \neq 0} \int_{\RR^n}  e^{-tL}\mu(dv) \frac{\left\vert e^{-2\pi i v.\xi} -1  + 2\pi i v.\xi \right\vert}{\vert \xi\vert^2} \leq \frac{(2\pi)^2}{2}\int_{\RR^n} e^{-tL}[\mu(dv)] \frac{(v.\xi)^2}{\vert \xi\vert^2}\,dv.
\end{equation}
\vspace{0.5cm}

This is of order $1$ after time of $O(\ln(n))$  by Lemma \ref{lm:quad}. We now study the second term $(1- R(v))/\vert \xi\vert^2$. It equals

\begin{eqnarray*}
     \frac{1}{\vert \xi\vert^2} \frac{\int [1- \cos( 2\pi \vert v\vert \vert \xi\vert \cos\theta_1)] (\sin\theta_1)^{n-2}\,d\theta_1}{ \int_0^{\pi} (\sin\theta_1)^{n-2}\, d\theta_1} &=& \frac{1}{\vert \vxi \vert^2} \frac{\int 2 \sin^2(\pi \vert v \vert \vert \xi\vert \cos\theta_1) (\sin\theta_1)^{n-2}\,d\theta_1}{ \int_0^{\pi} (\sin\theta_1)^{n-2}\, d\theta_1}\\ & \leq &  2\frac{\pi^2 \vert v\vert^2 \vert \xi\vert^2}{\vert \vxi\vert^2} \frac{\int_0^{\pi} \cos\theta_1^2\sin \theta_1^{n-2}\,d\theta_1}{\int_0^{\pi} \sin \theta_1^{n-2}\,d\theta_1}\\
           & = & \frac{(2\pi)^2 \vert v\vert^2}{2 n}.
\end{eqnarray*}

\noindent This, together with Lemma \ref{lm:quad},  proves the inequality in \eqref{eq:symmetric}.

\vspace{0.5cm}

To prove the inequality in \eqref{eq:propd2vcompare}, we need a way to ``liberate" $e^{-tL}$  so that Lemma $\ref{lm:d2}$ can be used. For $d_2(e^{-tL}\mu, e^{-tL}\nu)$, we can write

\begin{eqnarray*}
\vert \xi\vert^{-2} \left\vert e^{-tL} \hat{\mu}(\xi) - e^{-tL} \hat{\nu}(\xi) \right\vert &= &   \vert \xi\vert^{-2} \left\vert \int_{\RR^n}  e^{2\pi iv.\xi} \left( e^{-tL}\mu (dv) - e^{-tL}\nu(dv)\right) \right\vert \\ 
& = &  \vert \xi\vert^{-2} \left\vert \int_{\RR^n} [(e^{2\pi i v.\xi} - 1 + 2\pi i v.\xi) ]\left( e^{-tL}\mu (dv) - e^{-tL}\nu(dv)\right) \right\vert \\& \leq &  \frac{(2\pi)^2}{2} \vert \xi\vert^{-2}  \int_{\RR^n} (v.\xi)^2 \left\vert e^{-tL}[\mu (dv) - \nu(dv) ] \right\vert, \\
\end{eqnarray*}

\noindent as in inequality \eqref{eq:d2a1}. We now look at the term $\left\vert e^{-tL}[\mu (dv) - \nu(dv) ] \right\vert$. Let $A$ be a measurable set. Recall that $Q_{i,j}(\theta)[A] =\{ v: Q_{i,j}(\theta)[v] \in A\}$; $Q_{i,j}(\theta)$ can act on measures by the  adjoint action $[Q_{i,j}(\theta) \mu](A) := \mu [ Q_{i,j}(-\theta)[A]]$. We have

\begin{eqnarray*}
	\vert [Q_{i,j}\mu] (A) - [Q_{i,j}\nu](A)\vert & = & \left\vert \dashint_{0}^{2\pi} \left( [Q_{i,j}(\theta)\mu](A) -   [Q_{i,j}(\theta)\nu](A) \right)  d\theta \right\vert\\
	& = &  \left\vert \dashint_{0}^{2\pi} \left(\mu [Q_{i,j}(-\theta)(A)] -   \nu[Q_{i,j}(-\theta)(A)] \right)  d\theta \right\vert\\
	& \leq & \dashint_0^{2\pi} \left\vert \mu - \nu \right\vert (Q_{i,j}(-\theta)[A]) \, d\theta\\
	& = & Q_{i,j}\vert \mu - \nu \vert [A]
\end{eqnarray*}  
 
\noindent From the convexity of $s\mapsto \vert s \vert$  it follows that $\vert e^{-tL}[\mu (dv) - \nu(dv) ] \vert \leq e^{-tL}\vert \mu(dv) - \nu(dv)\vert$. Thus, we can use the self adjointness of $L$ and let $e^{-tL}$ act on $(v.\xi)^2$. 
This allows us to apply Lemma \ref{lm:quad} and obtain the desired upper bounds related to the second moment as follows.

\begin{eqnarray*}
\vert \vxi\vert^{-2} \left\vert e^{-tL} \hat{\mu}(\xi) - e^{-tL} \hat{\nu}(\xi) \right\vert & \leq &  \frac{(2\pi)^2}{2} \vert \vxi\vert^{-2}  \int_{\RR^n} e^{-tL} (v.\xi)^2 \left\vert \mu (dv) - \nu(dv)  \right\vert \\
 & \leq &  \frac{(2\pi)^2}{2} ((n-1) e^{-t} +1) \int_{\RR^n} \frac{\vert v\vert^2}{n}\left\vert \mu (dv) - \nu(dv)  \right\vert \hfill \square
\end{eqnarray*}

We cannot rule out the possibility that $d_2(\mu, R_{\mu})$ can be of order $n$ at $t=0$. In fact, if $\mu$ is a measure which is even, symmetric in its variables, and satisfies $\int \vec{v}\mu(dv)=\vec{0}, \int \vert v\vert^2\mu(dv)<\infty$, but $\int_{\RR^n} v_1 v_2 \mu(dv) \neq 0$, then we have

\begin{eqnarray*}
 	d_2( \mu, R_\mu) & \geq & \lim_{s\rightarrow 0} \lim_{\vxi = s(1,1,\dots, 1)} \vert \xi\vert^{-2} \left\vert \int \cos(2\pi v.\xi)  [\mu(dv) - R_\mu(dv)] \right\vert \\
                & = &  \lim_{s\rightarrow 0} \lim_{\vxi = s(1,1,\dots, 1)} \vert \xi\vert^{-2} \left\vert \int (\cos(2\pi v.\xi)-1)  [\mu(dv) - R_\mu(dv)] \right\vert \\
		& = &  \frac{(2\pi)^2}{2} \lim_{s\rightarrow 0} \lim_{\vxi = s(1,1,\dots, 1)}  \vert \xi\vert^{-2} \left\vert \int (v.\xi)^2  [\mu(dv) - R_\mu(dv)] \right\vert \\
		& = & \frac{(2\pi)^2}{2} \lim_{s\rightarrow 0} \lim_{\vxi = s(1,1,\dots, 1)}  \vert \xi\vert^{-2} \left\vert \int \left[ (v.\xi)^2- \frac{\vert v\vert^2\vert \xi\vert^2 }{n}\right] \mu(dv) \right\vert \\
		& = & \frac{(2\pi)^2}{2} \left\vert  \int v_1 v_2 \,\mu(dv) \right \vert  \lim\vert \xi\vert^{-2} \vert \sum_{i \neq j} \xi_i \xi_j \vert  = \frac{(n-1) (2\pi^2)}{2} \left\vert \int v_1 v_2 \mu(dv) \right\vert.
\end{eqnarray*}

\noindent and if $\mu$ is a even measure concentrated on the line $v_1 {=} v_2 {=} \dots {=} v_n$, then $\int v_1 v_2 \mu(dv) = \int v_1^2 \mu(dv) $ and $d_2(\mu, R_\mu)$ is a multiple of the total energy. Proposition \ref{prop:d2v} says that this condition won't last for time longer than $O(\ln(n))$. Also, if $\mu$ has mean zero and has all correlations zero (as in equation \eqref{eq:chaotic}) then $d_2(e^{-tL}\mu,R_{\mu})$  never becomes of order $n$ as shown by equation \eqref{eq:symmetric}.

\section{Proof of Theorem \ref{lm:d2}} \label{sec:upperbound} 
Let $\mu$ be a probability measure with mean zero and finite second moment and let $-L = n(Q-I)$. We use the fact that the Fourier transform commutes with the Kac evolution to take the problem into Fourier space. Because the second moment of $\mu$ is finite, $\hat{\mu}$ has bounded second derivatives. This will allow us to control $\vert \hat{\mu} - R_{\hat{\mu}}\vert_{L^\infty(r)}$ by $\vert  \hat{\mu} - R_{\hat{\mu}}\vert_{L^2(r)}^\frac{2}{n}$ on each sphere. The fact that the $L^2$ gap of the Kac operator in \cite{CCL} gives an exponential decay in $L^2(r)$ for each $r$ leads to a decay in $d_2(e^{-tL}\mu, R_\mu)$ after carefully obtaining order $r^{2}$ decay in $\left\vert e^{-tL}\hat{\mu}-\hat{R}_\mu \right\vert_{L^\infty(r)}$ as $r\rightarrow 0^+$ and combining the decay results on each sphere.

\vspace{0.5cm}

\noindent \textbf{Proof}(of Equation \eqref{eq:d2})  Let $u(t,\xi)$ be $\hat{\mu}(t,\xi) - \hat{R}_\mu(\xi)$. From equation  \eqref{eq:propd2vcompare} we have that

\begin{eqnarray}
 \left\vert \sum_{i,j} \eta_i \eta_j \partial_i \partial_j u(t,\xi) \right\vert & = &  \left\vert  -(2\pi)^2\int (\veta.\V)^2 e^{-2\pi i v.\xi} e^{-tL}(\mu - R_{\mu}) \right\vert    \nonumber\\
         & = &(2\pi)^2 \left\vert \int_{\RR^n} (\veta.\V)^2 e^{-2\pi i v.\xi} e^{-tL} \left\{ (I - R)[\mu]\right\} (dv)\right\vert  \nonumber \\
         & = &(2\pi)^2 \left\vert \int_{\RR^n} (\veta.\V)^2e^{-2\pi i v.\xi}  (I - R) [e^{-tL}(\mu)] (dv)   \right\vert \nonumber\\
         & = & (2\pi)^2 \left\vert \int_{\RR^n} (I - R)\left\{ (\veta.\V)^2e^{-2\pi i v.\xi} \right\} [e^{-tL}(\mu)] (dv)   \right\vert \nonumber \\
        & \leq & (2\pi)^2 \int  (I+R) (\veta.\V)^2  e^{-tL}[\mu](dv) = (2\pi)^2 \int e^{-tL} (\veta.\V)^2 \mu(dv)+ (2\pi)^2\frac{\vert v\vert^2 \vert \eta\vert^2}{n} \mu(dv) \nonumber  \\
       & = &(2\pi)^2 \vert \veta\vert^2 \left\{ 2 \int v_1^2 \mu(dv) + (n-1) e^{-\frac{4n-6}{n-1} t} \left\vert \int v_1 v_2 \mu(dv) \right\vert \right\} =:  L_p(t) \vert \veta\vert^2\label{eq:Laplace}
\end{eqnarray}

\noindent for all $\xi, \eta$ and all $t\geq 0$.\\

 Fix $t$ and $r>0$. Let $S = S^{n-1}(r)$ and  choose $\xi_0 \in S$ and $\theta_0$ so that $e^{-i\theta_0} u(\xi_0)= \vert u\vert_{L^\infty(S)}$. Let 

\[B = S \cap \left\{\vert \xi - \xi_0\vert \leq \sqrt{\frac{\vert u(\xi_0) \vert}{3Lp(t)}} \right\}.\]

All of $u, \xi_0$, and $B$ depend on $t$,  but we will suppress this dependence in many places. Our first task is to show that $\vert u\vert_{L^\infty(r)}$ is of order $r^2$ as $r\rightarrow 0$, for $d_2$ to be bounded. We will  accomplish this in equation \eqref{eq:claim2} which shows that  $\vert u(\xi_0)\vert -\vert u(\xi)\vert$ is actually quadratic in $\vert \xi-\xi_0\vert$ for $\xi \in B$. 

\vspace{\baselineskip}

Let us first show that $\vert u(\xi) \vert \geq \vert u(\xi_0)\vert/2$ on $B$. Let $\eta$ be any point in $\RR^n$. By Taylor's theorem we have:

\begin{equation}\label{eq:Taylor}
 u(\eta) = u(\xi_0)+ (\nabla u)(\xi_0).(\eta-\xi_0)+ \frac{1}{2}\sum_{i,j} \partial_{i}\partial_{j}u(\xi^\ast) (\eta-\xi_0)_i(\eta-\xi_0)_j.
\end{equation}

Equation \eqref{eq:Laplace} bounds the term $\frac{1}{2}\sum_{i,j} \partial_{i}\partial_{j}u(\xi^\ast) (\eta-\eta_0)_i(\eta-\eta_0)_j$  in absolute value by $\frac{1}{2} Lp(t) \vert \eta-\xi_0 \vert^2$. We next study the linear term in equation \eqref{eq:Taylor} when $\eta =\xi \in B$. 

\vspace{\baselineskip}

Since $\vert u(\xi)\vert^2$ has a maximum on $S$ at $\xi_0$, we have either $u(\xi_0)= 0$ or $\nabla \vert u(\xi_0)\vert$ is perpendicular to $S$ at $\xi_0$, and thus $\nabla u(\xi_0)$ is parallel to $\xi_0$. Without loss of generality we can take $u(\xi_0) \neq 0$ for otherwise $u\equiv 0$ on $S$ and $S$ does not contribute to $d_2$.

\vspace{\baselineskip}

If follows from our assumptions, including the assumption that $u(\xi_0)\neq 0$, that we have

\begin{equation}\label{eq:gradient}
 \vert \nabla u(\xi_0)\vert \leq Lp(t) \vert \xi_0\vert,
\end{equation}

\noindent which might be false at other points on $B$.\\

Equation \eqref{eq:gradient} follows from the following observations. First, the fact that $\nabla u(\xi_0)$ is parallel to $\xi_0$, 
thus $\displaystyle \vert \xi_0 . \nabla u(\xi_0) \vert = \vert \xi_0 \vert \vert \nabla(\xi_0)\vert$. Second, 

\[\displaystyle \vert \xi_0 . \nabla u(\xi_0) \vert= \left\vert \sum (\xi_0)_i \int_0^1 \partial_s (\partial_i u)(s \xi_0)\,ds\right\vert = 
\left\vert\sum_i \sum_j (\xi_0)_i (\xi_0)_j \int_0^1 \partial_j \partial_i u( s\xi_0)\,ds \right\vert \leq \vert \xi_0\vert^2 Lp(t)\]

\noindent by \eqref{eq:Laplace}. These observations prove equation \eqref{eq:gradient}.

\vspace{\baselineskip}

Note that equation \eqref{eq:Taylor} with $\xi_0$ replaced by zero imples that 

\begin{equation}\label{eq:promise}
 \vert u(\eta,t) \vert \leq \frac{Lp(t)}{2} \vert \eta \vert^2 \qquad \mbox{for any $\eta$ and $t$},
\end{equation}

\noindent since $u(t,0) = \int \mu(dv) - \int R_\mu(dv) = 0$ and $\nabla u(t,0) = -2\pi i \int \vec{v} e^{-tL}\mu(dv) - \vec{0} = \int e^{-nt} \vec{v} \mu(dv) = \vec{0}$ for all $t$.

\vspace{\baselineskip}

In particular, we have $\sqrt{\vert u(\xi_0) \vert} \leq \sqrt{\frac{Lp(t)}{2}} \vert \xi_0\vert$ and thus, for all $\xi \in B$ we have $\vert \xi - \xi_0\vert \leq \frac{1}{\sqrt{6}}\vert \xi_0\vert$. Hence $\xi.\xi_0 >0$ on $B$.

\vspace{\baselineskip} 

We  now find an upper bound for  $\vert \nabla u(\xi_0) . (\xi-\xi_0)\vert $ on S. We choose a coordinate system in which $\xi_0 = (0,\dots,0,0,r)$ and $\xi = (0, \dots, w, \sqrt{r^2 - w^2})$. Here we're using the fact that $\vec{\xi}.\vec{\xi_0}>0$ on $B$. Set the $n^{th}$ coordinate direction $\vec{e}_n$ to $\xi_0/r$. Then $\vert (\xi-\xi_0).e_n \vert = \vert r - \sqrt{r^2- w^2} \vert = \frac{w^2}{r+\sqrt{r^2-w^2}} \leq \frac{w^2}{r}$. Similarly, $\vert \xi-\xi_0 \vert^2 = w^2+(r-\sqrt{r^2-w^2})^2 = 2 r^2( 1 - \sqrt{1 - \frac{w^2}{r^2}}) \geq w^2$, which together with equation \eqref{eq:gradient}, gives the inequality 

\[  \vert (\nabla u)(\xi_0).(\xi-\xi_0)\vert \leq Lp(t) r \times \frac{w^2}{r} \leq Lp(t) \vert \xi-\xi_0\vert^2.\]

 In summary, we have shown that the for all $\xi \in  B$ the following inequality holds.

\begin{equation}\label{eq:claim2}
  \vert u(\xi_0) - u(\xi) \vert \leq \frac{3}{2} Lp(t) \vert \xi-\xi_0\vert^2.
\end{equation}

\noindent This implies that we have $\vert u(\xi) \vert \geq \vert u(\xi_0)\vert - \frac{3}{2}Lp(r)  \vert \xi-\xi_0\vert^2 \geq  \frac{\vert u(\xi_0)\vert}{2}$ on $B$. 

\vspace{\baselineskip}

 We complete the proof of equation \eqref{eq:d2} by a simple computation. We choose a coordinate system in which $\xi_0$ points towards the North Pole and we denote by $\theta$ the angle from the $\xi_0$ axis. The largest value $\theta_{max}$ of $\theta$ on $B$ satisfies the equation

\[\vert \xi - \xi_0\vert_{\mbox{max}}= 2 r \sin(\frac{1}{2}\theta_{\mbox{max}}).\]

By integrating out the rest of the angular variables in $\sigma^r$, we obtain

\begin{eqnarray*}
 \sigma^r(B) & = & \frac{ \int_0^{2\sin^{-1}\left(\sqrt{\frac{\vert u(\xi_0) \vert}{12 r^2\, Lp(t)}}\right)}
 \sin(\theta)^{n-2}\,d\theta}{\int_0^\pi \sin(\theta)^{n-2}\,d\theta} \geq \frac{ \int
 \sin(\theta)^{n-2} \cos(\theta)\,d\theta}{\int_0^\pi \sin(\theta)^{n-2}\,d\theta}\\ & = & \frac{ \left(4 \frac{\vert u(\xi_0) \vert}{12 r^2 Lp(t)}\left(1 -\frac{\vert u(\xi_0)\vert}{12 r^2 Lp(t)} \right)\right)^{(n-1)/2}}{(n-1) 
         \int_0^\pi \sin(\theta)^{n-2}\,d\theta} \geq \frac{ \left(\, \frac{23}{72} \frac{\vert u(\xi_0)\vert}{ Lp(t) r^2}\right)^{(n-1)/2}}{ (n-1) \int_0^\pi \sin(\theta)^{n-2}\,d\theta}.\\
\end{eqnarray*}

\noindent This gives us the lower bound $\displaystyle \vert \vert u(t,\xi) \vert\vert_{L^2(r)}^2  \geq \frac{\vert u(t,\xi_0)\vert^2}{4} \sigma^r(B)$. Letting $b(t,r)= \frac{\vert u(t, \xi_0)\vert}{r^2 Lp(t)}$, we obtain $b \leq \frac{1}{2}$ for all $t$ and have the following upper bound.

\begin{equation}\label{eq:16minus}
  \frac{\vert\vert u(t,.) \vert\vert_{L^2(r)}^2}{(Lp(t) r^2)^2} \leq e^{-2 \lambda_1 t} \frac{\vert\vert u(0,.) \vert\vert_{L^2(r)}^2}{(Lp(t) r^2)^2} \leq e^{-2\lambda_1 t} \frac{\vert\vert u(0,.)\vert\vert^2_{L^\infty(r)} }{(Lp(t) r^2)^2}\leq \left( \frac{Lp(0)}{Lp(t)}\right)^2\frac{e^{-2 \lambda_1 t}}{4}. 
\end{equation}

\noindent At the same time we have the following lower bound.

\begin{equation}\label{eq:16}
 \frac{\vert\vert u(t,.) \vert\vert_{L^2(r)}^2}{(Lp(t) r^2)^2} \geq \frac{\vert u(t,\xi_0)\vert^2}{4(Lp(t) r^2)^2} \sigma_r(B) \geq \frac{b(t,r)^2}{4}  \frac{ \left( \frac{23}{72} b(t,r) \right)^{(n-1)/2}}{ (n-1) \int_0^\pi \sin(\theta)^{n-2}\,d\theta}
\end{equation}

\noindent Equations \eqref{eq:16minus} and \eqref{eq:16} give the following inequality:

\[ b(t,r) \leq \frac{72}{23} e^{-\frac{4\lambda_1}{n+3}t} \left( (n-1)\left(\frac{n+1}{ne^{-t}+1}\right)^2 \int_0^\pi \sin(\theta)^{n-2}\,d\theta \times \left( \frac{72}{23} \right)^{\frac{n-1}{2}}\right)^{2/(n+3)}.\]

Finally,  since $n\geq 2$ and we have

\[\sup_{k\geq 2} \left( \frac{ 23^2}{72^2} \frac{(k-1) (k+1)^2}{(k e^{-t}+1)^2} \int_{0}^\pi \sin\theta^{k-2}\,d\theta \right)^{\frac{2}{k+3}} \leq 2.1207\]

\noindent (its the value when $k=6$ and $t=\infty$), we have $\displaystyle b(t,r)$ is less than or equal to  $\frac{72}{23}\times 2.1207 e^{-\frac{4\lambda_1}{n+3}}$ and $\displaystyle d_2(e^{-tL}\mu, R_{\mu})$ is at most $6.64 Lp(t) e^{-\frac{4\lambda_1}{n+3}t}$. $\hfill \square$ \\

\begin{rem} The proof of Theorem \ref{lm:d2} relies on equations \eqref{eq:16minus} and \eqref{eq:16} which can be seen as the norm $L^\infty$ being interpolated between $(L^{2})^{\frac{1}{n}}$ and $W^{2,\infty}$ \hspace{-0.1cm}.
$Lp(t)$ got through intact which potentially saves a factor $n$ compared to $Lp(0)$. It would be interesting function-analytically to see if more information than just $Lp(t)$ can be incorporated in this interpolation inequality using the exact form of $u$.
\end{rem}

\section{Construction of $f_0$}\label{sec:f0}
 In this section, for each $n\geq 2$ we construct a probability density $f_n$ on $\RR^n$ that is symmetric in its variables and has the property that 

\[ \frac{d_2( e^{-tL}f_n, R_{f_n})}{ d_2(f_n, R_{f_n})} \geq \max\{1 - \frac{e}{n} ( 2\lambda t)^{n-1}, 0\}. \]

\noindent This says that no matter how large $n$ is, $d_2(e^{-tL}f_n, R_{f_n})$ is practically unchanged for time at least $\frac{1}{2\lambda}$. Although this result provides no information about the decay after time of order $1$, it does rule out bounds of the form $d_2(f(t) ,R_f) \leq e^{-ct} d_2(f(0),R_f)$ for any $c$.
Let us rescale the time so that $\lambda=1$.

\vspace{\baselineskip}

In Lemmas $1{-}3$ we will construct a Schwartz function $\psi(v)$ for which 

\begin{equation}\label{eq:deaf}
d_2(Q^k \psi, R_\psi) = d_2(\psi, R_\psi) \mbox { for $k =0,1, \dots, n-2$.} 
\end{equation}

We will scale $\psi$ and add to it a positive Gaussian at large enough temperature to obtain a non-negative function $f_n$. The existence of $\psi$ satisfying equation \eqref{eq:deaf} is not very 
surprising and follows from the $L^\infty$ nature of the $d_2$ metric and the fact that it takes $n-1$ Kac rotations $Q$ of a vector $\vec{v}$ to cover the whole sphere 
$\vert \vec{w}\vert = \vert \vec{v}\vert$. This is analogous to the result in \cite{Diaconis} where it is shown that the total the variation distance between an
initial permutation of a deck of cards and the uniform distribution is not affected by $O(\ln(n))$ riffle-shuffles. The reason for this invariance is because there are permutations 
that cannot be reached in less than $O(\ln(n))$ riffle-shuffles.

\vspace{\baselineskip}
 
Since $d_2$ deals with the Fourier transforms, we will use the fact that the Fourier transform commutes with rotations, and thus with the Kac rotations $Q_{i,j}$ . We will directly construct the Fourier transform of the $f_n$-s and only afterwards ensure that the inverse Fourier transform is non-negative and in $L^1$. As a first step we will construct a one parameter family of functions $\phi(\xi;\alpha)\geq 0$ such that $\displaystyle Q^k \phi( (z, 0,0,\dots,0);\alpha) = 0$ for all $z,\alpha$ and all $k \leq n-2$.

\vspace{\baselineskip}

Let $h(x;\alpha) = (1 - e^{-\alpha x^2})$ and set $\phi(\xi; \alpha) = \prod_{i=1}^n h(\xi;\alpha)$ (We will drop the parameter $\alpha$ in $\phi$ below.). Then we have the following lemmas.

\begin{lem}{Properties of $\phi$ }\label{lm:basic}

 Fix $\vert \xi\vert=r$, and let $z_1 = (r,0,0,\dots,0)$. Then for all $l\leq n-2$ we have

\begin{enumerate}
 \item $[Q^l\phi](z_1)= \phi(z_1)=0$.
 \item $R_\phi(z_1) > \frac{1}{2}\vert \phi\vert_{L^\infty(r)}$; provided $\alpha\geq \alpha(r)$ is large enough.
 \item $\left[\frac{(nt)^{n-1}}{  (n-1)! }Q^{n-1}\phi\right] (z_1) \leq \frac{e}{n} (2t)^{n-1} \vert \phi \vert_{L^\infty(r)}$.
 \item $\vert \phi\vert_{L^\infty(r)} = (1 - e^{-\alpha r^2/n})^n$.
\end{enumerate}
 
\end{lem}

\begin{rem} Properties $(1)$ and $(3)$ are easier to prove for the function $\prod_{i=1}^n \xi_i^2$. We use $h(x;\alpha)$ instead of $x^2$ in $\phi$ to satisfy property $(2)$. Properties $(2)$ and $(1)$ tell us that the maximum of  
\[\left\vert \phi(\xi) - R_\phi \right\vert\]

\noindent on $S^{n-1}(r)$ is at $(\pm z_1, 0 ,\dots, 0)$ because we know that

\begin{equation*}
	R_\phi(r) -  \vert \phi\vert_{L^\infty(r)} \leq R_\phi(\xi) - \phi(\xi) \leq R_\phi(r),
\end{equation*}

\noindent and thus, on $S^{n-1}(r)$, we have

\begin{eqnarray*}
\left\vert \phi(\xi) - R_\phi \right\vert & \leq & \max\left\{ R_\phi(r), \left\vert R_\phi(r) - \vert \phi\vert_{L^\infty(r)}\right\vert \right\} = R_\phi(r)
\end{eqnarray*}

\noindent by property $(2)$.
\end{rem}

\begin{rem}
 The coefficient  of $Q^{n-1}[\phi](z_1)$ in property $(3)$ comes from the Taylor expansion of $e^{-nt Q} \phi$.
\end{rem}

\noindent \textbf{Proof} 
\begin{enumerate}
 \item Given  a sequence of Kac rotations $Q_{i_1, j_1}(\theta_1) , \dots, Q_{i_k,j_k} (\theta_k)$, we can define a sequence of trigonometric polynomials $\{ P_1^{(k)}, \dots, P_n^{(k)}\}_{k=1}^\infty$ as follows. Let 

\[ \left( \begin{array}{c} P_1^{(0)} \\ P_2^{(0)} \\ \dots \\ P_n^{(0)}\end{array}\right) = \left( \begin{array}{c} 1 \\ 0  \\ \dots \\ 0 \end{array}\right).\]

Once $\{ P_i^{(s)}\}_{i=1}^n$ are defined, define $P_i^{s+1}(\theta_1 , \dots, \theta_k)$ using the equality

\[ P_{i}^{(s+1)} = \left\{ \begin{array}{cl}  P_{i}^{(s)} (\theta_1,\dots, \theta_k), & i \not\in \{i_{s+1},j_{s+1}\}  \\
P_{i}^{(s)}(\theta_1,\dots, \theta_k) \cos(\theta_{s+1}) - P_{j_{s+1}}^{(s)} (\theta_1,\dots, \theta_k)\sin(\theta_{s+1}), & i=i_{s+1}\\
P_{i}^{(s)} (\theta_1,\dots, \theta_k) \sin(\theta_{s+1}) + P_{i_{s+1}}^{(s)}(\theta_1,\dots, \theta_k)\cos(\theta_{s+1}), & j=j_{s+1}  \end{array}\right. .\]

We are interested in these polynomials since they determine the velocity of particle $1$ after the $k$ Kac collisions above in the relation:
\[ v_1(\mbox{ after } ) = \sum_{i=1}^n P_i^{(k)}(\theta_1, \dots, \theta_k) v_i( \mbox{initial}).\]

We now show that if $i \geq 2$ is an index for which the ``edges" $\{ (i_1, j_1), \dots, (i_k,j_k)\}$ do not connect ``vertex" $i$ to vertex $1$, then $P_i(\theta_1, \dots,\theta_k)= 0$. 
Let $\mathcal{G}$ denote the graph on $(v_1, \dots, v_n)$ with edges $\{ (i_1, j_1), \dots, (i_k,j_k)\}$. Let $C$ be the connected component of $v_i$. An easy inductive argument
shows that $\{ P_j^{(l)} : j \in C\}$ depends only on $\{ P_j^{(0)}: j \in C\}$, for $l=0,1, \dots, k$.  In particular, $P_i^{(k)}$ is obtained from $\{ P_j^{(0)}(\theta_1, \dots, \theta_k): j \in C\}$
after possibly multiplying them by $\cos\theta$-s and $\sin\theta$-s, and adding them up. Since $P_j^{(0)} \equiv 0$ for $j \in C$, we have $P_i^{(k)}(\theta_1, \dots, \theta_k) \equiv 0$.

\vspace{\baselineskip}

As a conclusion, it follows that  if $ \left[ Q_{i_k, j_k} \dots Q_{i_1,j_1} h \right](z_1;\alpha) \neq 0$, then we have

\[ \left. Q_{i_k, j_k} \dots Q_{i_1,j_1} \prod \left(1 - e^{-\alpha \xi_i^2}\right) \right\vert_{z_1} = \frac{1}{(2\pi)^k} \int \prod_{i=1}^n \left(1 - e^{-\alpha r^2 (P_{i}^{(k)}(\{ \cos(\theta_l), \sin(\theta_l)\})^2}\right)   \prod_{j=1}^k d\theta_j \neq 0.\] 

Thus the connected component $C$ of $i$ must contain $1$ for each $i$. So $\mathcal{G}$ is a connected graph which means that $k\geq n-1$.   Property $(1)$ follows from the hypothesis that $k \leq n-2$.

\item For $r>0$ and $n\geq 2$ fixed, 
\[\frac{\phi}{\vert \phi \vert_{L^\infty(r)}} =  \frac{ \prod(1 - e^{-\alpha \xi_i^2})}{(1 - e^{-\alpha r^2/n})^n} \rightarrow 1\]

\noindent almost everywhere on $S^{n-1}(r)$ as ${\alpha r^2 \rightarrow \infty}$. Thus, by the dominated convergence theorem, there exists an $\bar{A}(n)<\infty$
such that if $\alpha r^2 \geq \bar{A}(n)$ then $\int_{S^{n-1}(r)} \phi(w) \sigma^r(dw) \geq \frac{1}{2} \vert \phi\vert_{L^\infty(r)}$. Let

\begin{equation}\label{eq:alphar}
\alpha(r,n) = \frac{\bar{A}(n)}{r^2}.
\end{equation}

Note that the property of having an $L^1(r)$ norm greater than or equal to $\frac{1}{2}$ the $L^\infty(r)$ norm is preserved in time under the Kac evolution $e^{-tL}$. This is because for positive functions, the Kac evolution does not change the $L^1$ norm, but it can only decrease the $L^\infty$ norm. This observation is also true when we replace $e^{-tL}$ by $Q^k$.

\item By Cayley's theorem there are $n^{n-2}$ distinct trees on $n$ vertices, and for each tree we can order its edges in $(n-1)!$ ways.  Each order of presentation of the edges in the tree comes with a weight ${n\choose 2}^{-(n-1)}$. The terms $Q_{i_{n-1}, j_{n-1}} \dots Q_{i_1, j_1}[\phi] (z_1)$ where the edges $\{ (i_1, j_1), \dots, (i_{n-1}, j_{n-1})\}$ do not connect all
the vertices $(v_1, \dots, v_n)$ evaluate to zero. The rest of the terms are non-negative and bounded above by $\vert \phi\vert_{L^\infty(r)}$. Thus,

\begin{equation}\label{eq:E4}
\frac{(nt)^{n-1}}{(n-1)!} (Q^{n-1}\phi )(z_1) \leq \frac{ (nt)^{n-1}}{(n-1)!} \frac{ (n-1)! n^{n-2}}{{n\choose 2}^{n-1}} \vert \phi\vert_{L^\infty(r)} \leq \frac{e}{n} (2t)^{n-1} \vert \phi \vert_{L^\infty(r)},
\end{equation}

proving property $(3)$.

\item This property follows from an application of the method of Lagrange multipliers. \hfill $\square$
\end{enumerate}

\noindent Since $\alpha(r,n)$ in the above lemma is proportional to $r^{-2}$, we need a way of keeping $r=\vert \xi\vert$ strictly away from zero when $d_2$ is being evaluated. We do this in Lemma \ref{lm:help} by multiplying. Let $\psi(\xi) = \phi(\xi)A(\xi)$, where $A(\xi) = \vert \xi\vert^4 e^{-\vert \xi\vert^2}$. Then we have the following Lemma.

\begin{lem}\label{lm:help}
Let $A(\xi) = \vert \xi\vert^4 e^{-\vert \xi\vert^2}$ and let $b$  be smallest solution to $(xe^{-x} = \frac{1}{2} e^{-1})$  ($b \approx 0.23196$). Let $\alpha =\alpha(\sqrt{b},n)$ be as in equation \eqref{eq:alphar}. If $\psi = A(\xi) \phi(\xi)$. Then $\frac{\vert \psi-R_\psi \vert}{\vert \xi\vert^2}$ has a maximum on $\RR^n - \{\vec{0}\}$ at a point $(x,0,0,\dots,0)$ with $x^2 \geq b$.
\end{lem}

\noindent \textbf{Proof} Choose $\alpha$ as in the hypothesis. Then $R_\phi(\xi) \geq \frac{1}{2}$  when $\vert \xi \vert \geq \sqrt{b}$ by property $2$ of Lemma \ref{lm:basic}. In particular:
$\frac{\vert \psi(1,0,\dots,0)-R_\psi(1)\vert}{\vert 1\vert^2} = e^{-b}\frac{R_\phi(1,0,0,\dots,0)}{1} \geq \frac{1}{2}e^{-b}\geq \frac{1}{2}e^{-1}$. So if $\vert \xi\vert^2 < b$,
then $\frac{\vert \psi(\xi)-R_\psi(\xi)\vert}{\vert \xi \vert^2} < b e^{-b} < \frac{1}{2}e^{-1}$. So we know that $the maximum \max\frac{\vert \psi-R_\psi \vert}{\vert \xi\vert^2} $ is attained at a point $\vec{\xi}$
 with norm at least $\sqrt{b}$. So, for our choice of $\alpha$, we have $R_\phi \geq \frac{1}{2} \vert\phi\vert_{L^\infty(r)}$ and property $1$ in Lemma \ref{lm:basic} shows that $\xi$ can be taken to be $(x,0,\dots,0)$ for some $x \geq \sqrt{b}$. $\hfill \square$

\vspace{\baselineskip}

We now give an explicit formula for $f_0$.

\begin{lem}\label{lm:f01}
  Let $b, \alpha=\alpha(\sqrt{b},n)$ be as in Lemma \ref{lm:help} and equation \eqref{eq:alphar}. Set

\[ f_0(v)  = \left( \frac{0.9 \pi}{1+\alpha}\right)^{\frac{n}{2}}  e^{- \left( \frac{0.9\pi^2}{1+\alpha}\right) \vert v\vert^2} +  \frac{1}{B (2\pi)^4} \triangle^2\prod_{i=1}^n \left( \sqrt{\pi} e^{-\pi^2 v_i^2} - \sqrt{\frac{\pi}{1+\alpha}} e^{- \frac{\pi^2}{1+\alpha} v_i^2}\right). \]

\noindent If $B>0$ is large enough, then $f_0$ is a probability density and equation \eqref{eq:slowdecay} holds for $f_0$.
\end{lem}

\noindent \textbf{Proof}
Notice that $f_0(v)$ is the sum of a Gaussian and $\frac{1}{B} \check{\psi}$. The Gaussian is radial at a high temperature since $\alpha$ is large. For large $\vert v_i\vert$, $\check{\psi}$ is bounded by polynomial of degree $4$ times $\exp(-\frac{\pi^2}{1+\alpha} \vert v\vert^2)$, so we can find a $B=B(n)$ that makes $\vert\check{\psi}\vert \leq B\left( \frac{0.9 \pi}{1+\alpha} \right)^{\frac{n}{2}}  e^{- \left( \frac{0.9 \pi^2}{1+\alpha}\right) \vert v\vert^2}$. This shows that when $B\geq B(n)$ we have $f_0 \geq 0$. Since $\psi$ is a Schwartz function, its Fourier transform is in $L^1$ and we have $\int \check{\psi}(v)\,dv = \psi(0)=0$. This shows that $f_0$ integrates to $1$.

\vspace{\baselineskip}

\noindent  We now prove equation \eqref{eq:slowdecay} for $f_0$. Note that $\frac{\left\vert e^{-tL}\hat{f_0}(\xi) - \hat{R}_{f_0}(\xi)\right\vert}{\vert \xi\vert^2} = \frac{\left\vert e^{-tL}\psi(\xi)- R_\psi \right\vert}{B\vert \xi\vert^2}$.
We showed in Proposition \ref{lm:help} that when $t=0$, this term is maximized at a point $z_1 = (z_0, 0,0,\dots,0)$ for some $z_0\geq \sqrt{b}$. Fix  $k\leq n-2$. Then

\begin{eqnarray}\label{eq:E3}
  0 & \geq& \frac{d_2(e^{-tL}f_0,R_{f_0}) - d_2(f_0,R_{f_0}) }{t^k} = \frac{d_2(e^{-tL}f_0,R_{f_0}) - \frac{R_\psi(z_1)}{B\,\vert z_1\vert ^2}}{ t^k} \nonumber\\
                           & \geq & \frac{1}{B t^k} \frac{(R_{\psi} (z_1) - e^{-tL}\psi)(z_1)) - R_\psi(z_1)}{z_0^2} = -\frac{ e^{-tL}\psi(z_1)}{B t^k z_0^2} = - \frac{ z_0^2 e^{-z_0^2}}{B t^k}  e^{-tL}\phi(z_1)
\end{eqnarray}

\noindent Here we used the fact that $e^{-tL}\psi$ and $\psi$ have the same radial parts.\\

\noindent Recall from Lemma \ref{lm:basic} that $Q^l\phi(z_1) = \phi(z_1) = 0$ for $l=0,1,2,\dots, n-2$.
 Hence, the same is true for their linear combinations $\left[ n^k(I-Q)^k \phi\right](z_1)$. Thus, by Taylor's theorem, the right hand side in equation \eqref{eq:E3} converges to $ -\frac{1}{z_0^2}\left(\frac{n^k}{n!}(I-Q)^k(\phi)(z_1)\right)$ as $t\rightarrow 0^+$,  which is zero if $k\leq n-2$. So $\left(e^{tnQ}\phi\right)(z_1) = \frac{n^{n-1}t^{n-1}}{(n-1)!} e^{t^\ast n Q} Q^{n-1}(\phi) (z_1)$ for some $t^\ast$ in $(0,t)$ and we have:

\begin{eqnarray*}
  0 & \geq & \frac{d_2(e^{-tL}f,R_f) - d_2(f,R_f) }{t^{n-1}}  = \frac{d_2(e^{-tL}f,R_f) - \frac{R_\psi(z_1)}{B\, z_0^2}}{t^{ n-1}}\\
      &  \geq &   - \frac{ z_0^2 e^{-z_0^2}}{B t^{n-1}}  e^{-tL}\phi(z_1) = -  \frac{ z_0^2 e^{-z_0^2}}{B} \frac{ n^{n-1}}{(n-1)!} \frac{e^{-nt} e^{t^\ast nQ} Q^{n-1}\phi (z_1)}{t^{n-1}}.\\
\end{eqnarray*}

\noindent Since $\left(e^{t^\ast nQ} Q^{ n-1 }\phi\right) (z_1)$ is less than $\vert Q^{n-1}\phi\vert_{L^\infty(z_0)} e^{t n} $,  we conclude that 

\[\frac{d_2(e^{-tL}f,R_f) - d_2(f,R_f) }{t^{n-1}}\geq -\frac{ n^{n-1}}{(n-1)!} z_0^2 e^{-z_0^2}\frac{\vert Q^{n-1} \phi\vert_{L^\infty(z_0)}}{B}.\] 

\noindent Combining this with property $(3)$ in Lemma \ref{lm:basic} gives equation \eqref{eq:slowdecay}.$\hfill \square$ 

\section{Proof of the Propagation of Chaos}\label{sec:chaos}

      McKean gave  in \cite{McKean} a short algebraic proof of propagation of chaos for Kac's original model 
on $S^{n-1}$. This proof was adapted in \cite{BLV} to give a propagation of chaos result for the fully thermostated Kac model. This section describes how McKean's proof can be further modified to give a propagation of chaos result for the partially thermostated Kac model in \cite{TV}.

\vspace{\baselineskip}

Let $Z=Z(\RR^\infty, \mbox{ symm })$ be the space of bounded and continuous functions depending on an arbitrary but finite but number of variables, endowed with the product 

\[f \otimes g(v_1, \dots, v_a, v_{a+1}, \dots, v_{a+b}) = \frac{1}{(a+b)!}\sum_\sigma  f(v_{\sigma(1)}, \dots, v_{\sigma(a)}) g(v_{\sigma(a+1)}, \dots, v_{\sigma(a+b)})\]

\noindent and identify functions which have the same symmetrization: $\displaystyle \int_{\RR^\infty} f\,\phi\,dv = \int_{\RR^\infty } g \, \phi\,dv$ for all $\phi \in L^1(\RR^\infty)$ that is symmetric in its variables. McKean observed that $n \lambda(Q -I)$ can be approximated by $2\lambda \Gamma$. Here $\Gamma$ is the operator given by 

\[ \Gamma[\phi(v_1, \dots, v_k)] = \sum_{i \leq k}  \dashint_{0}^{2\pi}\phi(v_1, \dots, v_i \cos\theta - v_{k+1} \sin\theta, v_{i+1}, \dots, v_k)) - \phi) \,d\theta, \]

\noindent that takes functions depending on $k$ variables to functions depending on $k+1$ variables. Note that $\Gamma$ is a derivation. That is, $\Gamma [f\otimes g] = \Gamma[f] \otimes g + f \otimes \Gamma[g]$. McKean demonstrated that propagation of chaos holds for $\{e^{tD}f_n\}_n$ whenever $D$ is a derivation. McKean then showed the terms in the Taylor expansion of $\int_{S^{n-1}}e^{t  \lambda n (Q-I)}f_n \phi d\sigma$ converged to the corresponding terms in $\int_{S^{n-1}} e^{2\lambda \Gamma}f_n \phi \,d\sigma$ as $n\rightarrow \infty$. Since both series converge absolutely when $t\propto \frac{1}{\lambda}$ is small enough, propagation of chaos follows.\\

The same proof was used in \cite{BLV} to show that there is  propagation of chaos for the fully thermostated Kac model. The observation there is that  the generator $-L = \eta \sum_{i=1}^n (M_i - I ) + n\lambda(Q-I)$ can be approximated by $\eta \sum_{i=1}^\infty ( M_i - I) + 2\lambda \Gamma$ which is a derivation. Here $M_i$ is the weaker Maxwellian thermostat acting on the $i^{th}$ particle: 
\[M_i[f] = \int_{\RR}\dashint_{0}^{2\pi} f(v_1, \dots, v_i \cos\theta-w\sin\theta, v_{i+1}, \dots, v_n) g(v_i \sin\theta + w \cos\theta)\,d\theta\,dw\] 

We will tweak this proof, which works on both $\{L\left( S^{n-1}\right)\}_n$ and $\{L^1\left( \RR^n\right)\}_n$, for the partially thermostated Kac model. Suppose $\alpha = \frac{m_0}{n_0}$ is the fraction of thermostated particles. Thermostating part of the particles divides the indices $1, \dots, n$ into two groups $A_n$ (the thermostated) and $B_n$ (the rest).  Our initial condition $f_n(0, .)$ should be symmetric under the exchange of particles in $A_n$ and under the exchange of particles in $B_n$. We want to have a space similar to $Z$ and a derivation similar to $\Gamma$ that adapt to the fact that a new particle introduced in the system is not always thermostated.

\vspace{\baselineskip}

 One approach is to let the underlying space be $\bar{Z}=\bar{Z}((\RR^{n_0})^\infty)$ and to let $f$, $g$ all depend on $k n_0$ , $l n_0$ variables.  We can let every particle with index $i \equiv 1,2, \dots, m_0 (\,mod \,n_0)$ to be thermostated. We can define $f\otimes g$  analogously by
\[f \otimes g(v_1, \dots, v_{k n_0}, v_{k n_0+1}, \dots, v_{(k +l)n_0}) = \frac{1}{((k+l) m_0)!((k+l) (n_0-m_0))!}\sum_\sigma  f(v_{\sigma(1)}, \dots, v_{\sigma(k n_0)}) g(v_{\sigma(k n_0+1)}, \dots, v_{\sigma((k+l)n_0)}).\]

\noindent Here $\sigma$ runs over all permutations leaving $A_n$ (and also $B_n$) invariant. Our generator becomes $-\mathcal{L}_k$ given by the equation
\[ -\mathcal{L}_k = k n_0 \lambda( Q-I) + \eta\sum_{i=1}^{k n_0} \mathbf{1}_{[1,\dots, m_0]} (i \mod \,n_0) (P_i - I).\]

We replace $\Gamma$ by $\bar{\Gamma}:\bar{Z}\mapsto \bar{Z}$ that takes functions depending on $k n_0$ variables to functions on $(k+1)n_0$ variables. $\bar{\Gamma}$ is given by
\[ \bar{\Gamma}[\phi](v_1, \dots, v_{(k+1)n_0} ) = \sum_{i \leq kn_0} \sum_{l= kn_0 + 1}^{(k+1)n_0} \dashint_{0}^{2\pi}\phi(v_1, \dots, v_i \cos\theta - v_{k+1} \sin\theta, v_{i+1}, \dots, v_k)) - \phi) \,d\theta. \]

We see that $\displaystyle \bar{\Gamma} = 2\lambda \Gamma +  \eta\sum_{i=1}^{k n_0} \mathbf{1}_{[1,\dots, m_0]} (i \mod \,n_0) (P_i - I)$. Hence $\bar{\Gamma}$ is a derivation. Note that we have the inequality

\[ \left\vert\left\vert \mathcal{L}_k\phi - 2 \lambda \bar{\Gamma}[\phi] - \eta\sum_{i=1}^{k n_0} \mathbf{1}_{[1,\dots, m_0]} (i \mod \,n_0) (P_i - I)[\phi]
 \right\vert\right\vert \leq \frac{l^2 n_0}{k} 4 { l n_0 \choose 2} \vert\vert\phi\vert\vert + 2\lambda \frac{l n_0}{k n_0+1} \vert\vert \bar{\Gamma} \phi\vert\vert, \]

\noindent  whenever $\phi$ depends only on $l n_0$ variables with $l<k$. This goes to $0$ when $l$ is fixed and $k \rightarrow \infty$.

\vspace{\baselineskip}

Finally, for every $k, l \geq 0$, we  have the following bound 

\[\vert\vert \mathcal{L}_{k+l} \circ \mathcal{L}_{k+l-1} \circ \dots \circ \mathcal{L}_{k+1} \circ \mathcal{L}_k f \vert\vert_\infty \leq (4\lambda + 2\eta)^{(l+1)} k (k+1) \dots (k+l-1)
 \vert\vert f\vert\vert_\infty.\]

\noindent This makes $\sum_l \frac{t^l}{l!} \vert\vert\mathcal{L}_{k+l-1} \circ \dots \circ \mathcal{L}_{k+1} \circ \mathcal{L}_k f \vert\vert_\infty$ converge for all $k$ when $t < \frac{1}{4\lambda+2\eta}$. McKean's proof can be used step by step from this point on (see also Lemma $19$ in \cite{BLV}) to give propagation of chaos for time $t=\frac{0.9}{4\lambda+2\eta}$. Iterating this process $j$-times shows propagation of chaos for time up to $\frac{0.9 j}{4\lambda+2\eta}$, and hence for all $t>0$ since $j$ is arbitrary.
 
\section{Conclusion}\label{sec:conc}
We saw in Theorem \ref{lm:d2} that under the Kac evolution a Borel measure $\mu$ approaches its angular average $R_\mu$ in the GTW metric $d_2$ exponentially with rate at least $O\left(\frac{1}{n}\right)$ and saw in Theorem \ref{Th:on} that the initial decay in $d_2$ can be very slow at least for time $1/(2\lambda)$ which is a macroscopic quantity.  We also saw that the average energy per particle also controls $d_2(\mu, R_\mu)$ after time of order $\ln(n)$.  Proposition \ref{prop:d2v} suggests that the constant $K$ in Theorem \ref{lm:d2} is not optimal. This raises the question of what is the optimal $K(n)$? And whether our conjecture in \eqref{eq:conjecture} is correct. The proof of Theorem \ref{lm:d2} gives an application of the $L^2$ gap to initial states that are not necessarily in $L^1(\RR^n) \cap L^2(\RR^n)$ and can be generalized to other evolutions which have gaps in $L^2$  provided their generators commute with the Fourier transform. For example: the Kac model in $1$ dimension with an initial state not symmetric in its variables; the Kac model in $1$ dimension with symmetric collision rules for which $\theta$ in \eqref{eq:Qij} has weight $\rho(\theta)$ where $\rho$ is
not necessarily constant but satisfies $\rho(2\pi-\theta)= \rho(\theta)$. It would be interesting to check if decay rates for Fourier based metrics can be obtained for non-Maxwellian molecules, where the collision rate between particles $i$ and $j$ is proportional to $\vert v_i^2 +v_j^2 \vert^{\frac{\gamma}{2}}$ for some $\gamma$ in $(0,2]$; or for the momentum conserving Kac model in $3$ dimensions with Maxwellian molecules whose gap was computed in 
$\cite{CGL}$. The functions $\{f_n\}$ suggest a set of questions such as: can there be a sequence of distributions $\mu_n$ similar to the $\{f_n\}$-s except that they are supported on the sphere? and, since the $f_n$ are small $L^1$-perturbations of Gaussians  by Schwartz functions with a very particular algebraic structure ,  is there a physical interpretation to these structures? or, can we find functions $\tilde{f}_n$ similar to the $f_n$ for which there is a physical interpretation? Our lower bound in Theorem \ref{Th:on} is effective only when $t\leq \frac{1}{2\lambda}$. It should be possible to make this bound effective for a longer time interval by improving the upper bound in property (3) of Lemma \ref{lm:basic}. If we improve the bound $\left\vert  Q_{i_{n-1},j_{n-1}}\dots Q_{i_2,j_2} Q_{i_1,j_1}\phi](z_1)] \right\vert \leq \left\vert \phi\right\vert_{L^\infty(r)}$ in equation \eqref{eq:E4}, we will have a larger lower bound for $d_2(e^{-tL}\mu, R_\mu)$.  


\section{Acknowledgement} I am grateful to Federico Bonetto and Michael Loss for fruitful discussions and thank Ranjini Vaidyanathan for helping  me with Theorem \ref{Th:chaos}. I was partially supported by the NSF grant DMS-$1301555$. I also thank
the anonymous referee for helpful feedback and suggestions.

\end{document}